\documentstyle[12pt,epsf,amsfonts,amssymb,amsbsy]{article}
\newcommand\VV{\setbox0=\hbox{V}\hbox{\rm V\raise\ht0
  \hbox to0pt{\hss\vbox to0pt{\hbox{v}\vss}}}}
\def\slashchar#1{\setbox0=\hbox{$#1$}           
   \dimen0=\wd0                                 
   \setbox1=\hbox{/} \dimen1=\wd1               
   \ifdim\dimen0>\dimen1                        
      \rlap{\hbox to \dimen0{\hfil/\hfil}}      
      #1                                        
   \else                                        
      \rlap{\hbox to \dimen1{\hfil$#1$\hfil}}   
      /                                         
   \fi}                                         %

\begin{document}
\vspace*{3cm}
\begin{center}
{\large \bf Lifetimes of $\Xi_{bc}^{+}$ and
$\Xi_{bc}^{0}$ baryons. }\\
\vspace*{5mm}
V.V. Kiselev, A.K. Likhoded, A.I. Onishchenko\\
{\sf State Research Center of Russia "Institute for High Energy Physics"} \\
{\it Protvino, Moscow region, 142284 Russia}\\
Fax: +7-095-2302337\\
E-mail: kiselev@mx.ihep.su
\end{center}
\begin{abstract}
Estimates of lifetimes and partial branching ratios for the baryons 
$\Xi_{bc}^{+}$ and
$\Xi_{bc}^{0}$ are presented using the inverse heavy quark mass expansion 
technique carried out in the Operator Product Expansion approach.
We take into account both the  perturbative QCD corrections to the
spectator contributions and, depending on the
quark contents of hadrons, the Pauli interference and weak scattering
effects between the constituents, using a  potential model  for the 
evaluation of the non-perturbative parameters.
\end{abstract}

\section{Introduction}
Recently, techniques based on the inverse heavy quark mass $(1/m_Q)$ 
expansion
have been applied to study the properties of heavy hadrons in QCD 
\cite{HQET,NRQCD,BB}. These techniques are embedded in the  Operator 
Product Expansion (OPE) approach used in the context of effective theories.
They allow to calculate non-perturbative effects in the decays of
heavy hadrons in terms of a few universal quantities, enabling to extract 
the parameters of the standard model,
such as the weak mixing angles involving quarks and heavy quark masses. 
The attained theoretical accuracy, reflecting the convergence of the 
series in both the inverse heavy quark mass and the QCD coupling constant, 
can be systematically improved and allows to make precise predictions in the 
heavy quark sector in the standard model (SM), such as decay rates and 
distributions and partial rate asymmetries involving {\sf CP} 
violation \footnote{For review see \cite{al}.}. The
virtual corrections due to "new" physics at a much higher scale
may influence the decay characteristics of the heavy hadrons,
enabling to measure deviations from 
the SM-predictions in existing and forthcoming experiments.

The approach under discussion has been convincingly used in the
description of weak decays of the hadrons having a single heavy quark,
carried out in the
framework of the heavy quark effective theory (HQET) \cite{HQET}, in the 
annihilation and radiative decays of
heavy quarkonia $Q\bar Q$, using
the framework of non-relativistic QCD (NRQCD) \cite{NRQCD}, and in the weak 
decays of long-lived heavy
quarkonium with mixed heavy flavours $B_c^+$ \cite{BB} \footnote{The first
experimental observation of the $B_c$-meson has been reported by the CDF 
Collaboration
\cite{CDFbc}; see \cite{thbc} for a review of the theoretical status of 
the $B_c$-meson decays.}.  In particular, we note that the
experimental data on the weak decays of hadrons with two heavy quarks can be
used to determine the parameters entering in the theoretical 
description of these systems. In turn, these parameters
could also be determined in well-defined frameworks, say, the 
potential model approach.

The baryons with two heavy quarks, $QQ^\prime q$, provide a new insight in the
description of systems containing the heavy quarks. For these baryons we can
apply a method, based on the combined HQET-NRQCD  techniques 
\cite{HQET,NRQCD,BB}, if we use the quark-diquark picture for the bound 
states. The expansion in the inverse heavy quark mass for the heavy 
diquark $QQ^\prime$ is a straightforward generalization of these 
techniques in the  mesonic decays \cite{NRQCD,BB}, with the difference 
that one is dealing, instead of the  colour-singlet systems, with the  
colour--anti-triplet ones, with the appropriate account of
the interaction with the light quark.
First estimates of the lifetimes for the doubly charmed
baryons $\Xi_{cc}$ have been recently presented in \cite{ltcc}. The
spectroscopic characteristics of baryons with two heavy quarks and the
mechanisms of their production in different interactions have been 
discussed in refs.~\cite{spec} and \cite{prod}, respectively.

In this paper, we present the calculation of
lifetimes for the baryons $\Xi_{bc}^{+}$ and $\Xi_{bc}^{0}$, containing
the beauty and charmed quarks. As in the description of inclusive decays
of the $\Xi_{cc}$-baryons, we follow the papers \cite{vs,BB} where the
necessary generalizations to the case of hadrons with two heavy
quarks and other corrections are discussed. We note, that keeping just the
leading term in the OPE, the inclusive widths are  determined by the 
mechanism of
spectator decays involving free quarks, wherein the corrections in the
perturbative QCD are taken into account. The inclusion of subleading terms in
the expansion in the inverse heavy quarks mass\footnote{It was shown in
\cite{bigi} that the leading order correction in $1/m_Q$ is absent and the
corrections begin with $1/m_Q^2$..} allows one to take into account the
corrections due to the quark confinement inside the hadron. In this way,
an essential role is played by the following non-perturbative 
characteristics:
 the motion of heavy quark inside the hadron and the
corresponding time dilation in the hadron rest frame with respect to the
quark rest frame, and the influence of the chromomagnetic interaction of 
the quarks. The
important ingredient of such corrections in the baryons with two heavy quarks
is the presence of a compact heavy diquark.
 The next peculiarity of baryons with two heavy quarks
is the significant numerical influence on the lifetimes by the quark
contents of hadrons, since in the third order in the inverse heavy quark
mass, $1/m_Q^3$,  the four-quark correlations in the total width are 
enforced in the
effective lagrangian due to the two-particle phase space in the intermediate
state (see the discussion in \cite{vs}). In this situation, we have to add
the effects of the Pauli interference between the products of heavy quark
decays and the quarks in the initial state as well as the
scattering involving the quarks composing the hadron. Through such
terms we introduce the corrections involving the masses of light and strange
quarks in the framework of non-relativistic models with the constituent quarks.
We include the
corrections to the effective weak lagrangian due to the evolution of
Wilson coefficients from the scale of the order of heavy quark
mass to the energy, characterizing the binding of quarks inside the hadron.

This paper is organized as follows. 
Following the picture given above, we describe the general scheme for the
construction of OPE for the total width of baryons with two heavy quarks with
account of the corrections to the spectator width in Section 2. In Section 3,
the procedure for the estimation of non-perturbative matrix elements of the
states of doubly heavy baryons is considered for the operators of
non-relativistic heavy quarks. Section 4 is devoted to the numerical evaluation
of lifetimes for $\Xi_{bc}^{+}$ and $\Xi_{bc}^{0}$ and partial decay 
rates, as well as to the discussion
of the underlying uncertainties. We conclude in section 5 by summarizing 
our results.

\section{Operator Product Expansion for the heavy baryons}

In accordance with the optical theorem, the total width
$\Gamma_{\Xi_{cc}^{\diamond}}$ for the baryon $\Xi_{bc}^{\diamond}$ with
$\diamond$, denoting the electrical charge, has the form
\begin{equation}
\Gamma_{\Xi_{bc}^{\diamond}} =
\frac{1}{2M_{\Xi_{bc}^{\diamond}}}\langle\Xi_{bc}^{\diamond}|{\cal T}
|\Xi_{bc}^{\diamond}\rangle ,
\label{1}
\end{equation}
where we accept the ordinary relativistic normalization of state, $\langle
\Xi_{bc}^{\diamond}|\Xi_{bc}^{\diamond}\rangle  = 2EV$, and the transition
operator ${\cal T}$:
\begin{equation}
{\cal T} = \Im m\int d^4x~\{{\hat T}H_{eff}(x)H_{eff}(0)\},
\end{equation}
is determined by the effective lagrangian of weak interaction $H_{eff}$ at the
characteristic hadron energies:
\begin{equation}
H_{eff} = \frac{G_F}{2\sqrt 2}V_{q_2q_3}V_{Qq_1}^{*}[C_{+}(\mu)O_{+} +
C_{-}(\mu)O_{-}] + h.c. 
\end{equation}
where
$$
O_{\pm} = [\bar q_{1\alpha}\gamma_{\nu}(1-\gamma_5)Q_{\beta}][\bar
q_{2\gamma}\gamma^{\nu}(1-\gamma_5)q_{3\delta}](\delta_{\alpha\beta}\delta_{
\gamma\delta}\pm\delta_{\alpha\delta}\delta_{\gamma\beta}),
$$
$$
C_+ = \left [\frac{\alpha_s(M_W)}{\alpha_s(\mu)}\right ]^{\frac{6}{33-2f}},
\quad
C_- = \left [\frac{\alpha_s(M_W)}{\alpha_s(\mu)}\right ]^{\frac{-12}{33-2f}},\\
$$
so that $f$ denotes the number of flavors, and $Q$ marks the flavor of heavy
quark ($b$ or $c$).       

The quantity ${\cal T}$ in (\ref{1}) permits the Operator Product Expansion
in the inverse powers of heavy quark mass, which determines the energy
release in the inclusive decays of baryon, containing the heavy quarks. Then
the total width $\Gamma_{\Xi_{bc}^{\diamond}}$ contains the series in the
non-perturbative matrix elements of operators with the increasing dimensions
in energy. In this way, for the subsystem, composed of the heavy quarks
only, we can additionally use the smallness of relative velocity $v$ in the
movement of heavy quarks in contrast to the interaction with the light quark,
wherein the only small-order parameter is the ratio of non-perturbative scale to
the heavy quark mass $\frac{\Lambda_{QCD}}{m_Q}$. Thus, there is the additional
energy scale in the heavy subsystem along with the parameter given by the
ratio of quark-gluon condensate scale to the heavy quark mass. It is the
relative momentum of quarks $m_Q v$, which is much less than the energy release
in the heavy quark decay, too.

So, OPE has the form:
\begin{equation}
{\cal T} = \sum_{i=1}^2\{C_1(\mu)\bar Q^iQ^i +
\frac{1}{m_{Q^i}^{2}}C_2(\mu)\bar
Q^ig\sigma_{\mu\nu}G^{\mu\nu}Q^i
+ \frac{1}{m_{Q^i}^{3}}O(1)\}. \label{4}
\end{equation}

The leading contribution is given by the spectator decay, i.e. by the term 
$\bar QQ$, which is the operator of dimension 3. The equations of 
motion allow to
show that there is no contribution by the operators with the dimension 4.
Further, there is only operator of dimension 5: $Q_{GQ} = \bar Q g
\sigma_{\mu\nu} G^{\mu\nu} Q$.  Among the operators of dimension 6, $Q_{2Q2q} =
\bar Q\Gamma q\bar q\Gamma^{'}Q$, the dominant contributions are provided by
the Pauli interference and weak scattering. The latter ones are enforced by the
two-particle phase space in comparison to the operators $Q_{61Q} = \bar Q
\sigma_{\mu\nu} \gamma_{l} D^{\mu}G^{\nu l}Q$ and $Q_{62Q} = \bar Q D_{\mu}
G^{\mu\nu} \Gamma_{\nu}Q$, which are neglected in what follows.

Then we have
\begin{eqnarray}
{\cal T}_{\Xi_{bc}^{+}} &=& {\cal T}_{35b} + {\cal T}_{35c} + 
{\cal T}_{6,PI}^{(1)} + {\cal T}_{6,WS}^{(1)},\nonumber\\
{\cal T}_{\Xi_{bc}^{0}} &=& {\cal T}_{35b} + {\cal T}_{35c} + 
{\cal T}_{6,PI}^{(2)} + {\cal T}_{6,WS}^{(2)},
\nonumber
\end{eqnarray}
where two initial terms denote the contributions into the decays of quark $Q$
by the operators with the dimensions 3 and 5, and the forthcoming terms are the
interference and scattering of constituents. In the explicit form we find
\begin{equation}
{\cal T}_{35b} = \Gamma_{b,spec}\bar bb - \frac{\Gamma_{0b}}{m_b^2}[2P_{c1} +
P_{c\tau 1} + K_{0b}(P_{c1} + P_{cc1}) + K_{2b}(P_{c2} + P_{cc2})]O_{Gb},
\label{5}
\end{equation}
\begin{equation}
{\cal T}_{35c} = \Gamma_{c,spec}\bar cc - \frac{\Gamma_{0c}}{m_c^2}[(2 +
K_{0c})P_{s1} + K_{2c}P_{s2}]O_{Gc}, \label{6}
\end{equation}
where
\begin{equation}
\Gamma_{0b} = \frac{G_F^2m_b^5}{192{\pi}^3}|V_{cb}|^2\qquad , \Gamma_{0c} =
\frac{G_F^2m_c^5}{192{\pi}^3}
\end{equation}
with $K_{0Q} = C_{-}^2 + 2C_{+}^2,~K_{2Q} = 2(C_{+}^2 - C_{-}^2)$ and  
$\Gamma_{Q,spec}$ denotes the spectator width (see \cite{bigi,9,10,11}):
\begin{equation}
P_{c1} = (1-y)^4,\quad P_{c2} = (1-y)^3,
\end{equation}
\begin{eqnarray}
P_{c\tau 1} &=& \sqrt{1-2(r+y)+(r-y)^2}[1 - 3(r+y) + 3(r^2+y^2) - r^3 - y^3
-4ry + \nonumber\\
&& 7ry(r+y)] + 12r^2y^2\ln\frac{(1-r-y+\sqrt{1-2(r+y)+(r-y)^2})^2}{4ry}
\end{eqnarray}
\begin{equation}
P_{cc1} = \sqrt{1-4y}(1 - 6y + 2y^2 + 12y^3)
24y^4\ln\frac{1+\sqrt{1-4y}}{1-\sqrt{1-4y}}
\end{equation}
\begin{equation}
P_{cc2} = \sqrt{1-4y}(1 + \frac{y}{2} + 3y^2)
- 3y(1-2y^2)\ln\frac{1+\sqrt{1-4y}}{1-\sqrt{1-4y}}
\end{equation}
\noindent
where $y = \frac{m_c^2}{m_b^2}$ and $r = m_{\tau}^2/m_b^2$. The functions
$P_{s1} (P_{s2})$ can be obtained from $P_{c1} (P_{c2})$ by the
substitution $y = m_s^2/m_c^2$. In the $b$-quark decays, we neglect the
value $m_s^2/m_b^2$ and suppose $m_s = 0$. 

The calculation of both the Pauli interference effect for the products of 
heavy quark
decays with the quarks in the initial state and the weak scattering of quarks,
composing the hadron, results in:
\begin{eqnarray}
{\cal T}_{6,PI}^{(1)} &=& {\cal T}_{PI,u\bar d}^c + {\cal T}_{PI,s\bar c}^b +
{\cal T}_{PI,d\bar u}^b + \sum_l{\cal T}_{PI,l\bar\nu_l}^b\\
{\cal T}_{6,PI}^{(2)} &=& {\cal T}_{PI,s\bar c}^b +
{\cal T}_{PI,d\bar u}^b + {\cal T}_{PI,d\bar u}^{'b} + \sum_l{\cal
T}_{PI,l\bar\nu_l}^b\\
{\cal T}_{6,WS}^{(1)} &=& {\cal T}_{WS,bu} +
{\cal T}_{WS,bc}\\
{\cal T}_{6,WS}^{(2)} &=& {\cal T}_{WS,cd} +
{\cal T}_{WS,bc}
\end{eqnarray}
\noindent
so that
\begin{eqnarray}
{\cal T}_{PI,s\bar c}^b &=&
-\frac{G_F^2|V_{cb}|^2}{4\pi}m_b^2(1-\frac{m_c}{m_b})^2\nonumber\\
&& ([(\frac{(1-z_{-})^2}{2}- \frac{(1-z_{-})^3}{4}) 
(\bar b_i\gamma_{\alpha}(1-\gamma_5)b_i)(\bar
c_j\gamma^{\alpha}(1-\gamma_5)c_j) + \nonumber\\ 
&& (\frac{(1-z_{-})^2}{2} -
\frac{(1-z_{-})^3}{3})(\bar b_i\gamma_{\alpha}\gamma_5 b_i)(\bar
c_j\gamma^{\alpha}(1-\gamma_5)c_j)] \label{16}
\\&& [(C_{+} - C_{-})^2 + 
\frac{1}{3}(1-k^{\frac{1}{2}})(5C_{+}^2+C_{-}^2+6C_{-}C_{+})]+ \nonumber\\
&& [(\frac{(1-z_{-})^2}{2} - \frac{(1-z_{-})^3}{4})(\bar
b_i\gamma_{\alpha}(1-\gamma_5)b_j)(\bar c_j\gamma^{\alpha}(1-\gamma_5)c_i) +
\nonumber\\
&&  (\frac{(1-z_{-})^2}{2} - \frac{(1-z_{-})^3}{3})(\bar
b_i\gamma_{\alpha}\gamma_5b_j)(\bar
c_j\gamma^{\alpha}(1-\gamma_5)c_i)]k^{\frac{1}{2}}(5C_{+}^2+C_{-}^2+6C_{-}C_{+}
)),\nonumber\\
{\cal T}_{PI,\tau\bar\nu_{\tau}}^b &=&
-\frac{G_F^2|V_{cb}|^2}{\pi}m_b^2(1-\frac{m_c}{m_b})^2\nonumber\\
&& [(\frac{(1-z_{\tau})^2}{2} - \frac{(1-z_{\tau})^3}{4})(\bar
b_i\gamma_{\alpha}(1-\gamma_5)b_j)(\bar c_j\gamma^{\alpha}(1-\gamma_5)c_i) +
\label{17}\\
&&  (\frac{(1-z_{\tau})^2}{2} - \frac{(1-z_{\tau})^3}{3})(\bar
b_i\gamma_{\alpha}\gamma_5b_j)(\bar
c_j\gamma^{\alpha}(1-\gamma_5)c_i)],\nonumber\\
{\cal T}_{PI,d\bar u}^{b'} &=&
-\frac{G_F^2|V_{cb}|^2}{4\pi}m_b^2(1-\frac{m_d}{m_b})^2\nonumber\\
&& ([(\frac{(1-z_{-})^2}{2}- \frac{(1-z_{-})^3}{4}) 
(\bar b_i\gamma_{\alpha}(1-\gamma_5)b_i)(\bar
d_j\gamma^{\alpha}(1-\gamma_5)d_j) + \nonumber\\ 
&& (\frac{(1-z_{-})^2}{2} -
\frac{(1-z_{-})^3}{3})(\bar b_i\gamma_{\alpha}\gamma_5 b_i)(\bar
d_j\gamma^{\alpha}(1-\gamma_5)d_j)] \label{18}
\\&& [(C_{+} + C_{-})^2 + 
\frac{1}{3}(1-k^{\frac{1}{2}})(5C_{+}^2+C_{-}^2-6C_{-}C_{+})]+ \nonumber\\
&& [(\frac{(1-z_{-})^2}{2} - \frac{(1-z_{-})^3}{4})(\bar
b_i\gamma_{\alpha}(1-\gamma_5)b_j)(\bar d_j\gamma^{\alpha}(1-\gamma_5)d_i) +
\nonumber\\
&&  (\frac{(1-z_{-})^2}{2} - \frac{(1-z_{-})^3}{3})(\bar
b_i\gamma_{\alpha}\gamma_5b_j)(\bar
d_j\gamma^{\alpha}(1-\gamma_5)d_i)]k^{\frac{1}{2}}(5C_{+}^2+C_{-}^2-6C_{-}C_{+}
)),\nonumber\\
{\cal T}_{PI,u\bar d}^c &=&
-\frac{G_F^2}{4\pi}m_c^2(1-\frac{m_u}{m_c})^2\nonumber\\
&& ([(\frac{(1-z_{-})^2}{2}- \frac{(1-z_{-})^3}{4}) 
(\bar c_i\gamma_{\alpha}(1-\gamma_5)c_i)(\bar
u_j\gamma^{\alpha}(1-\gamma_5)u_j) + \nonumber\\ 
&& (\frac{(1-z_{-})^2}{2} -
\frac{(1-z_{-})^3}{3})(\bar c_i\gamma_{\alpha}\gamma_5 c_i)(\bar
u_j\gamma^{\alpha}(1-\gamma_5)u_j)] \label{19}
\\&& [(C_{+} + C_{-})^2 + 
\frac{1}{3}(1-k^{\frac{1}{2}})(5C_{+}^2+C_{-}^2-6C_{-}C_{+})]+ \nonumber\\
&& [(\frac{(1-z_{-})^2}{2} - \frac{(1-z_{-})^3}{4})(\bar
c_i\gamma_{\alpha}(1-\gamma_5)c_j)(\bar u_j\gamma^{\alpha}(1-\gamma_5)u_i) +
\nonumber\\
&&  (\frac{(1-z_{-})^2}{2} - \frac{(1-z_{-})^3}{3})(\bar
c_i\gamma_{\alpha}\gamma_5c_j)(\bar
u_j\gamma^{\alpha}(1-\gamma_5)u_i)]k^{\frac{1}{2}}(5C_{+}^2+C_{-}^2-6C_{-}C_{+}
)),\nonumber\\
{\cal T}_{WS,bc} &=&
\frac{G_F^2|V_{cb}|^2}{4\pi}m_b^2(1+\frac{m_c}{m_b})^2(1-z_{+})^2[(C_{+}^2 +
C_{-}^2 +
 \frac{1}{3}(1 - k^{\frac{1}{2}})(C_{+}^2 - C_{-}^2))\nonumber\\
&&(\bar b_i\gamma_{\alpha}(1
- \gamma_5)b_i)(\bar c_j\gamma^{\alpha}(1 - \gamma_5)c_j) + \label{20}\\
&& k^{\frac{1}{2}}(C_{+}^2 - C_{-}^2)(\bar b_i\gamma_{\alpha}(1 - \gamma_5)b_j)
(\bar c_j\gamma^{\alpha}(1 - \gamma_5)c_i)],\nonumber\\
{\cal T}_{WS,bu} &=&
\frac{G_F^2|V_{cb}|^2}{4\pi}m_b^2(1+\frac{m_u}{m_b})^2(1-z_{+})^2[(C_{+}^2 +
C_{-}^2 +
 \frac{1}{3}(1 - k^{\frac{1}{2}})(C_{+}^2 - C_{-}^2))\nonumber\\
&&(\bar b_i\gamma_{\alpha}(1
- \gamma_5)b_i)(\bar u_j\gamma^{\alpha}(1 - \gamma_5)u_j) + \label{21}\\
&& k^{\frac{1}{2}}(C_{+}^2 - C_{-}^2)(\bar b_i\gamma_{\alpha}(1 - \gamma_5)b_j)
(\bar u_j\gamma^{\alpha}(1 - \gamma_5)u_i)],\nonumber\\
{\cal T}_{WS,cd} &=&
\frac{G_F^2}{4\pi}m_c^2(1+\frac{m_d}{m_c})^2(1-z_{+})^2[(C_{+}^2 + C_{-}^2 +
 \frac{1}{3}(1 - k^{\frac{1}{2}})(C_{+}^2 - C_{-}^2))\nonumber\\
&&(\bar c_i\gamma_{\alpha}(1
- \gamma_5)c_i)(\bar d_j\gamma^{\alpha}(1 - \gamma_5)d_j) + \label{22}\\
&& k^{\frac{1}{2}}(C_{+}^2 - C_{-}^2)(\bar c_i\gamma_{\alpha}(1 - \gamma_5)c_j)
(\bar d_j\gamma^{\alpha}(1 - \gamma_5)d_i)],\nonumber 
\end{eqnarray}
\begin{eqnarray}
{\cal T}_{PI,d\bar u}^b &=& {\cal T}_{PI,s\bar c}^b~(z_{-}\to 0)\\
{\cal T}_{PI,e\bar\nu_e}^b &=& {\cal T}_{PI,\mu\bar\nu_{\mu}}^b = 
{\cal T}_{PI,\tau\bar\nu_{\tau}}^b~(z_{\tau}\to 0),
\end{eqnarray}
where the following notation has been used:
\begin{eqnarray}
\mbox{in Eq.}~(\ref{16})&& z_{-} = \frac{m_c^2}{(m_b-m_c)^2},\quad k =
\frac{\alpha_s(\mu)}{\alpha_s(m_b-m_c)},\nonumber\\
\mbox{in Eq.}~(\ref{17})&& z_{\tau} = 
\frac{m_{\tau}^2}{(m_b-m_c)^2},\quad k =
\frac{\alpha_s(\mu)}{\alpha_s(m_b-m_c)},\nonumber\\
\mbox{in Eq.}~(\ref{18})&& z_{-} = \frac{m_c^2}{(m_b-m_d)^2},\quad k =
\frac{\alpha_s(\mu)}{\alpha_s(m_b-m_d)},\nonumber\\
\mbox{in Eq.}~(\ref{19})&& z_{-} = \frac{m_s^2}{(m_c-m_u)^2},\quad k =
\frac{\alpha_s(\mu)}{\alpha_s(m_c-m_u)},\nonumber\\
\mbox{in Eq.}~(\ref{20})&& z_{+} = \frac{m_c^2}{(m_b+m_c)^2},\quad k =
\frac{\alpha_s(\mu)}{\alpha_s(m_b+m_c)},\nonumber\\
\mbox{in Eq.}~(\ref{21})&& z_{+} = \frac{m_c^2}{(m_b+m_u)^2},\quad k =
\frac{\alpha_s(\mu)}{\alpha_s(m_b+m_u)},\nonumber\\
\mbox{in Eq.}~(\ref{22})&& z_{+} = \frac{m_s^2}{(m_c+m_d)^2},\quad k =
\frac{\alpha_s(\mu)}{\alpha_s(m_c+m_d)}.\nonumber
\end{eqnarray}
In the evolution of coefficients $C_{+}$ and
$C_{-}$, we have taken into account the threshold effects, connected to the
heavy quark masses.

In expressions (\ref{5}) and  (\ref{6}), the scale $\mu$ has been taken
approximately equal to $m_c$. In the Pauli interference term, we suggest 
that the scale can be determined on the basis of the agreement of the
experimentally known difference between the lifetimes of $\Lambda_c$,
$\Xi_c^{+}$ and $\Xi_c^{0}$ with the theoretical predictions in the framework
described above\footnote{A more expanded description is presented
in \cite{ltcc}.}. In any case, the choice of the normalization scale leads 
to 
uncertainties in the final results. Theoretical accuracy can be improved 
by the
calculation of next-order corrections in the powers of QCD coupling constant.

The contributions by the leading terms $\bar bb$ and $\bar cc$, as they 
stand in
(\ref{4}), correspond to the imaginary parts of the diagrams like the one 
shown in
Fig.~1. The coefficients for these operators are equivalent to the widths 
of the decays of
free quarks and are known in the logarithmic approximation of QCD to the 
second
order \cite{12,13,14,15,16}, including the mass corrections in the final state
with the charmed quark and $\tau$-lepton \cite{16} in the decays of $b$-quark
and with the strange quark mass for the decays of the $c$-quark. To 
calculate the corrections
to the logarithmic approximation, it is necessary to know the Wilson
coefficients in the effective weak lagrangian to the next-to-leading order and
the corrections with the single-gluon exchange in Fig.~1. In the
numerical estimates, we include these corrections and mass effects, but we
neglect the decay modes suppressed by the Cabibbo angle, and also the 
strange quark mass effects in $b$ decays. The bulky expressions for 
the spectator widths are given in the Appendix of \cite{ltcc}.

The $\sum_{i=1}^2O_{GQ^i}$ contributions are obtained by the calculation of
diagrams like one in Fig.~1 with all possible insertions of external gluon
attached to the internal quark lines. The corresponding expressions are known
in the logarithmic approximation. Finally, the contributions by the operators
with dimension 6 are obtained by cutting a single internal quark
line in Fig.~1, which results in diagrams of Fig.~2. These terms
corresponds to the Pauli interference of decay products with the quarks from
the initial state, and the weak scattering of quark composing the hadron. We
have calculated these effects with account of the masses in the final
states
and for the logarithmic renormalization of the effective lagrangian
for the non-relativistic heavy quarks at energies less than the heavy
quark masses.

To calculate the semileptonic decay branching fraction for the  
baryon
$\Xi_{bc}^{\diamond}$, we have used the expressions from \cite{10,16}:
\begin{eqnarray}
\Gamma_{sl} &=& \sum_{i=1}^2
4\Gamma_{Q^i}(\{1-8\rho^i+8\rho^{i3}-\rho^{i4}-12\rho^{i2}\ln\rho^i\}
+\nonumber\\
&&
E_{Q^i}\{5-24\rho^i+24\rho^{i2}-8\rho^{i3}+3\rho^{i4}-12\rho^{i2}\ln\rho^i\}+\\
&&
K_{Q^i}\{-6+32\rho^i-24\rho^{i2}-2\rho^{i4}+24\rho^{i2}\ln\rho^i\}+\nonumber\\
&& G_Q^i\{-2+16\rho^i-16\rho^{i3}+2\rho^{i4}+24\rho^{i2}\ln\rho^i\}),\nonumber 
\end{eqnarray}
where
\begin{equation}
\Gamma_c = |V_{cs}|^2G_F^2\frac{m_c^5}{192\pi^3},\qquad \Gamma_b =
|V_{cb}|^2G_F^2\frac{m_b^5}{192\pi^3},
\end{equation}
\begin{equation}
\rho^1=\frac{m_s^2}{m_c^2},\qquad \rho^2=\frac{m_c^2}{m_b^2}.
\end{equation}
The quantities $E_{Q^i} = K_{Q^i} + G_{Q^i}$, $K_{Q^i}$ and $G_{Q^i}$ are
determined by the formulae:
\begin{eqnarray}
K_{Q^i} &=& -\langle \Xi_{bc}^{\diamond}(v)|\bar
Q_v^i\frac{(iD)^2}{2m_{Q^i}^2}Q_v^i|\Xi_{bc}^{\diamond}(v)\rangle ,\nonumber\\
G_{Q^i} &=& \langle \Xi_{bc}^{\diamond}(v)|\bar
Q_v^i\frac{gG_{\alpha\beta}\sigma^{\alpha\beta}}{4m_{Q^i}^2}Q_v^i|\Xi_{bc}^{(*)
}(v)
\rangle ,
\end{eqnarray}
whereas the spinor field of effective theory $Q_v^i$ is given by the form:
\begin{equation}
Q^i(x) = e^{-im_{Q^i}v\cdot
x}\Bigl[1+\frac{i\slashchar{D}}{m_{Q^i}}\Bigr]Q_v^i(x).
\end{equation}
We have also taken into account the known $\alpha_s$-corrections to the
semileptonic quark decay width.

Thus, the calculation of the lifetime for the baryon $\Xi_{bc}^{\diamond}$ is
reduced to the problem of evaluation of the matrix elements of the 
operators, which is the the subject of the next section.

\section{Hadronic matrix elements}
In accordance with the equations of motion, the matrix element of operator 
$\bar Q^jQ^j$ can be expanded in series in powers of ${1}/{m_{Q^j}}$:
\begin{eqnarray}
\langle \Xi_{bc}^{\diamond}|\bar Q^jQ^j|\Xi_{bc}^{\diamond}\rangle _{norm} = 1
-
\frac{\langle \Xi_{bc}^{\diamond}|\bar
Q^j[(i\boldsymbol{D})^2-(\frac{i}{2}\sigma
G)]Q^j|\Xi_{bc}^{\diamond}\rangle_{norm}}{2m_{Q^j}^2} +
O(\frac{1}{m_{Q^j}^3}).
\end{eqnarray}
So, we have to estimate the numerical values involving  
the following set of operators:
\begin{eqnarray}
&& \bar Q^j(i\boldsymbol{ D})^2Q^j,\quad (\frac{i}{2})\bar Q^j\sigma GQ^j,\quad
\bar
Q^j\gamma_{\alpha}(1-\gamma_5)Q^j\bar q\gamma^{\alpha}(1-\gamma_5)q,\nonumber\\
&& \bar Q^j\gamma_{\alpha}\gamma_5Q^j\bar q\gamma^{\alpha}(1-\gamma_5)q,\quad 
\bar Q^j\gamma_{\alpha}\gamma_5Q^j\bar
Q^k\gamma^{\alpha}(1-\gamma_5)Q^k,\\ 
&& \bar Q^j\gamma_{\alpha}(1-\gamma_5)Q^j\bar
Q^k\gamma^{\alpha}(1-\gamma_5)Q^k.\nonumber
\end{eqnarray}
The first term corresponds to the motion of the quark inside the hadron, 
and it
results in the corrections caused by the time dilation in the hadron rest
frame with respect to the quark rest frame. The 
second term introduces the
corrections due to the chromomagnetic interaction of quarks. The third and
fourth terms depend on the four-quark fields and are connected with 
the contributions from  the Pauli interference and weak scattering.

Further, following the general approach of effective theories, we introduce the
effective field $\Psi_Q$, which, for the case under consideration, represents
the nonrelativistic spinor of heavy quark, so that we account for the
virtualities $\mu$ in the rage of $m_{Q} >  \mu >  m_{Q}v_{Q}$ in the framework
of perturbative QCD. The nonperturbative effects in the matrix elements have to
be expressed in terms of effective non-relativistic fields. So, we have
\begin{eqnarray}
\bar QQ &=& \Psi_Q^{\dagger}\Psi_Q -
\frac{1}{2m_Q^2}\Psi_Q^{\dagger}(i\boldsymbol{
D})^2\Psi_Q +
\frac{3}{8m_Q^4}\Psi_Q^{\dagger}(i\boldsymbol{ D})^4\Psi_Q -\nonumber\\
&& \frac{1}{2m_Q^2}\Psi_Q^{\dagger}g\boldsymbol{\sigma}\boldsymbol{ B}\Psi_Q -
\frac{1}{4m_Q^3}\Psi_Q^{\dagger}(\boldsymbol{ D}g\boldsymbol{ E})\Psi_Q + ...
\label{32}\\
\bar Qg\sigma_{\mu\nu}G^{\mu\nu}Q &=&
-2\Psi_Q^{\dagger}g\boldsymbol{\sigma}\boldsymbol{ B}\Psi_Q -
\frac{1}{m_Q}\Psi_Q^{\dagger}(\boldsymbol{ D}g\boldsymbol{ E})\Psi_Q + ...
\label{33}
\end{eqnarray}
We omit the term of $\Psi_Q^{\dagger}\boldsymbol{\sigma} (g\boldsymbol{E}
\times \boldsymbol{ D})\Psi_Q$, corresponding to the spin-orbital interactions,
which are not essential for the basic state of baryons under consideration. For
the normalization, we suppose 
\begin{equation}
\int d^3x\Psi_Q^{\dagger}\Psi_Q = \int d^3x Q^{\dagger}Q.
\end{equation} 
Then, for $Q$ defined as
\begin{equation}
Q\equiv e^{-imt}\left(\phi\atop \chi\right)~,
\end{equation}
we find
\begin{equation}
\Psi_c = \left( 1 + \frac{(i\boldsymbol{ D})^2}{8m_c^2}\right)\phi~.
\end{equation}

Let us stress the difference between the descriptions for the interaction of
a heavy quark with the light one and of a heavy quark with the heavy one. 
As we
have already mentioned above, in the heavy subsystem there is an additional
parameter which is the relative velocity of the quark. It introduces the
energy scale equal to $m_Q v$. Therefore, the Darwin term ($\boldsymbol{
D}\boldsymbol{ E}$) in the heavy subsystem stands  in the same order in the
inverse heavy quark mass in comparison to the chromomagnetic term
($\boldsymbol{\sigma}\boldsymbol{ B}$) (they have the same power in the
velocity $v$). This statement becomes evident if we apply the scaling
rules in NRQCD \cite{4}:
$$
\Psi_Q\sim (m_Qv_Q)^{\frac{3}{2}},\quad|\boldsymbol{ D}|\sim m_Qv_Q,\quad
gE\sim m_Q^2v_Q^3,\quad gB\sim m_Q^2v_Q^4,\quad g\sim v_Q^{\frac{1}{2}}.
$$
For the interaction of heavy quark with the light one, there is no such small
velocity parameter, so that the Darwin term is suppressed by the additional
factor of $k/m_Q\sim \Lambda_{QCD}/m_Q$.

Further, the experience of phenomenology in the potential quark models shows,
that the kinetic energy of quarks practically does not depend on the quark
contents of system, and it is determined by the color structure of state. So,
we suppose that the kinetic energy is equal to $T = m_dv_d^2/2 + m_lv_l^2/2$
for the quark-diquark system, and it is $T/2 = m_{b}v_{b}^2/2 +
m_{c}v_{c}^2/2$ in the diquark case (note the color factor of 1/2). Then
\begin{equation}
\frac{\langle
\Xi_{bc}^{\diamond}|\Psi_c^{\dagger}(i\boldsymbol{D})^2\Psi_c|\Xi_{bc}^{
\diamond}
\rangle }{2M_{\Xi_{bc}^{\diamond}}m_c^2}\simeq
v_c^2\simeq
\frac{2m_lT}{(m_l+m_b+m_c)(m_b+m_c)}+\frac{m_bT}{m_c(m_c+m_b)}.
\end{equation}
\begin{equation}
\frac{\langle
\Xi_{bc}^{\diamond}|\Psi_b^{\dagger}(i\boldsymbol{D})^2\Psi_b|\Xi_{bc}^{
\diamond}
\rangle }{2M_{\Xi_{bc}^{\diamond}}m_b^2}\simeq
v_b^2\simeq
\frac{2m_lT}{(m_l+m_b+m_c)(m_b+m_c)}+\frac{m_cT}{m_b(m_c+m_b)}.
\end{equation}
Numerically, we assign $T\simeq 0.4$ GeV, which results in $v_c^2 =0.195$ and
$v_b^2 =0.024$, where the dominant contribution is provided by the motion
inside the diquark.

Define:
\begin{eqnarray}
O_{mag} &\equiv& \frac{g_s}{4m_c}\bar c\sigma_{\mu\nu}G^{\mu\nu}c +
\frac{g_s}{4m_b}\bar b\sigma_{\mu\nu}G^{\mu\nu}b,\\
O_{mag} &=& \frac{\lambda}{m_c}(S_{cl}(S_{cl}+1) - S_c(S_c+1) - 
S_l(S_l+1)) +\\
&& \frac{\lambda}{m_b}(S_{bl}(S_{bl}+1) - S_b(S_b+1) - S_l(S_l+1)),\nonumber
\end{eqnarray}
where $S_{bl} = S_b + S_l$, $S_{cl} = S_c + S_l$ , $S_b$ is the $b$-quark spin,
$S_c$ is that of $c$-quark, and $S_l$ is the light quark spin. The operator
under study is related to the hyperfine splitting in the baryon system:
\begin{eqnarray}
&& \langle S_{bc} = 1, S = \frac{3}{2}|O_{mag}|S_{bc} = 1, S =
\frac{3}{2}\rangle
- \langle S_{bc} = 1, S = \frac{1}{2}|O_{mag}|S_{bc} = 1, S =
\frac{1}{2}\rangle~~~\nonumber\\
&& = \langle S_{bc} = 1, S = \frac{3}{2}|V_{hf}|S_{bc} = 1, S =
\frac{3}{2}\rangle
- \langle S_{bc} = 1, S = \frac{1}{2}|V_{hf}|S_{bc} = 1, S =
\frac{1}{2}\rangle,~~~
\end{eqnarray}
so that $S$ denotes the total spin of the system, and $S_{bc}$ is the 
diquark spin.
Further, the perturbative term, depending on the spins, is equal to
\begin{equation}
V_{hf} = \frac{8}{9}\alpha_s\frac{1}{m_lm_c}{\bf S}_l{\bf S}_c|R^{dl}(0)|^2 + 
\frac{8}{9}\alpha_s\frac{1}{m_lm_b}{\bf S}_l{\bf S}_b|R^{dl}(0)|^2,
\end{equation}
where $R^{dl}(0)$ is the radial wave function at the origin of quark-diquark
system. In contrast to the diquark system with the identical quarks, this
operator is not diagonal in the basis of $S$ and $S_{bc}$. To proceed, we use:
\begin{equation}
|S;S_{bc}\rangle  = \sum_{S_{bl}} (-1)^{(S+S_l+S_c+S_b)}\sqrt
{(2S_{bl}+1)(2S_{bc}+1)}
\left\{\begin{array}{ccc} \bar S_l & S_b & S_{bl}\label{44} \\
                        S_c & S & S_{bc} \end{array}\right\}|S;S_{bl}\rangle 
\end{equation}
and 
\begin{equation}
|S;S_{bc}\rangle  = \sum_{S_{cl}} (-1)^{(S+S_l+S_c+S_b)}\sqrt
{(2S_{cl}+1)(2S_{bc}+1)}
\left\{\begin{array}{ccc} \bar S_l & S_c & S_{cl}\label{45} \\
                        S_b & S & S_{bc} \end{array}\right\}|S;S_{cl}\rangle 
\end{equation}
The result of substitutions gives
\begin{equation}
\lambda = \frac{4|R^{dl}(0)|^2\alpha_s}{9m_l},
\end{equation}
which can be used in the calculations for the baryon with the vector
diquark, however, in what follows we put $S_{bc} = 0$, so that
\begin{equation}
\frac{\langle\Xi_{bc}^{\diamond}|O_{mag}|\Xi_{bc}^{\diamond}\rangle}{2M_{\Xi_{b
c}^{\diamond}}}
= 0
\end{equation}
The account of Darwin and chromomagnetic terms results in:
\begin{eqnarray}
\frac{\langle \Xi_{bc}^{\diamond}|\Psi_c^{\dagger}g\boldsymbol{\sigma}\cdot
\boldsymbol{B}\Psi_c|\Xi_{bc}^{\diamond}\rangle }{2M_{\Xi_{bc}^{\diamond}}} &=&
-\frac{2}{3}g^2\frac{|\Psi^d(0)|^2}{m_b}, \\
\frac{\langle \Xi_{bc}^{\diamond}|\Psi_c^{\dagger}(\boldsymbol{ D}\cdot
g\boldsymbol{E})\Psi_c|\Xi_{bc}^{\diamond}\rangle }{2M_{\Xi_{bc}^{\diamond}}}
&=&
\frac{2}{3}g^2|\Psi^d (0)|^2.
\end{eqnarray}
where $\Psi^d (0)$ is the wave function at the origin of diquark. The analogous
matrix elements\footnote{The obtained expressions differ from those for the
$B_c$ meson \cite{BB} because of the color structure of the state, providing
the factor of $\frac{1}{2}$.} for the operators of beauty quarks can be written
down by the permutation of heavy quark masses.

Combining the results above, we find
\begin{eqnarray}
\frac{\langle \Xi_{bc}^{\diamond}|\bar cc|\Xi_{bc}^{\diamond}\rangle
}{2M_{\Xi_{bc}^{\diamond}}} &=& 1 -
\frac{1}{2}v_c^2 + \frac{g^2}{3m_bm_c^2}|\Psi^d (0)|^2 - \nonumber\\
&& \frac{1}{6m_c^3}g^2|\Psi^d (0)|^2 + \ldots \nonumber\\
&\approx& 1 - 0.097 +0.004 - 0.007\ldots
\end{eqnarray}
The dominant role in the corrections is played by the term, connected to the
time dilation because of the quark motion inside the diquark. Next, for the
operator $cg\sigma_{\mu\nu}G^{\mu\nu}c$ we have
\begin{eqnarray}
\frac{\langle \Xi_{bc}^{\diamond}|\bar
cg\sigma_{\mu\nu}G^{\mu\nu}c|\Xi_{bc}^{\diamond}\rangle
}{2M_{\Xi_{bc}^{\diamond}}m_c^2} &=&
\frac{4g^2}{3m_b m_c^2}|\Psi^d (0)|^2 - 
\frac{g^2}{3m_c^3}|\Psi^d (0)|^2 \approx 0.002.
\end{eqnarray}
The permutation of quark masses lead to the required expressions for the
operators of $\bar bb$ and $\bar bg\sigma_{\mu\nu}G^{\mu\nu}b$.

For the four quark operators, determining the Pauli interference and the weak
scattering, we use the estimates in the framework of non-relativistic potential
model \cite{ltcc}:
\begin{eqnarray}
(\bar b\gamma_{\mu}(1-\gamma_5)b)(\bar c\gamma^{\mu}(1-\gamma_5)c) &=&
2(m_c+m_b)|\Psi^{d} (0)|^2(1-4S_bS_c),\\
(\bar b\gamma_{\mu}\gamma_5b)(\bar c\gamma^{\mu}(1-\gamma_5)c) &=&
-4S_bS_c\cdot 2(m_c+m_b)|\Psi^{d} (0)|^2,\\
(\bar b\gamma_{\mu}(1-\gamma_5)b)(\bar q\gamma^{\mu}(1-\gamma_5)q) &=&
2(m_b+m_l)|\Psi^{dl} (0)|^2(1-4S_bS_q),\\
(\bar b\gamma_{\mu}\gamma_5b)(\bar q\gamma^{\mu}(1-\gamma_5)q) &=&
-4S_bS_q\cdot 2(m_b+m_l)|\Psi^{dl} (0)|^2,\\
(\bar c\gamma_{\mu}(1-\gamma_5)c)(\bar q\gamma^{\mu}(1-\gamma_5)q) &=&
2(m_c+m_l)|\Psi^{dl} (0)|^2(1-4S_cS_q),\\
(\bar c\gamma_{\mu}\gamma_5c)(\bar q\gamma^{\mu}(1-\gamma_5)q) &=&
-4S_cS_q\cdot 2(m_c+m_l)|\Psi^{dl} (0)|^2.
\end{eqnarray}
The exploitation of (\ref{44}) and (\ref{45}) for the basic states of baryons
results in
\begin{eqnarray}
\langle \Xi_{bc}^{\diamond}|(\bar b\gamma_{\mu}(1-\gamma_5)b)(\bar
c\gamma^{\mu}(1-\gamma_5)c)|\Xi_{bc}^{\diamond}\rangle  &=& 8(m_b+m_c)\cdot
|\Psi^{d}(0)|^2,\\
\langle \Xi_{bc}^{\diamond}|(\bar b\gamma_{\mu}\gamma_5b)(\bar
c\gamma^{\mu}(1-\gamma_5)c)|\Xi_{bc}^{\diamond}\rangle  &=& 6(m_b+m_c)\cdot
|\Psi^{d}(0)|^2,\\
\langle \Xi_{bc}^{\diamond}|(\bar b\gamma_{\mu}(1-\gamma_5)b)(\bar
q\gamma^{\mu}(1-\gamma_5)q)|\Xi_{bc}^{\diamond}\rangle  &=& 2(m_b+m_l)\cdot
|\Psi^{dl}(0)|^2,\\
\langle \Xi_{bc}^{\diamond}|(\bar b\gamma_{\mu}\gamma_5b)(\bar
q\gamma^{\mu}(1-\gamma_5)q)|\Xi_{bc}^{\diamond}\rangle  &=& 0,\\
\langle \Xi_{bc}^{\diamond}|(\bar c\gamma_{\mu}(1-\gamma_5)c)(\bar
q\gamma^{\mu}(1-\gamma_5)q)|\Xi_{bc}^{\diamond}\rangle  &=& 2(m_c+m_l)\cdot
|\Psi^{dl}(0)|^2,\\
\langle \Xi_{bc}^{\diamond}|(\bar c\gamma_{\mu}\gamma_5c)(\bar
q\gamma^{\mu}(1-\gamma_5)q)|\Xi_{bc}^{\diamond}\rangle  &=& 0.
\end{eqnarray}
The color structure of wave functions leads to the relations
$$\langle \Xi_{bc}^{\diamond}|(\bar
c_iT_{\mu}c_k)(\bar q_k\gamma^{\mu}(1-\gamma_5)q_i)|\Xi_{bc}^{\diamond}\rangle
=
-\langle \Xi_{bc}^{\diamond}|(\bar cT_{\mu}c)(\bar
q\gamma^{\mu}(1-\gamma_5)q)|\Xi_{bc}^{\diamond}\rangle ,
$$
where $T_{\mu}$ is an arbitrary spinor matrix.

\section{Numerical estimates}

Calculating the inclusive widths of decays for the $\Xi_{bc}^{+}$ and
$\Xi_{bc}^{0}$ baryons, we have used the following set of dimensional
parameters in the model:
\begin{equation}
\begin{array}{lcllcl}
m_c & = & 1.6, & m_l & = & 0.3, \\
m_b & = & m_c+3.5, & m_s & = & m_l+0.15, \\
\mu & = & 1.2, & T & = & 0.4, 
\end{array}
\end{equation}
where all numbers are in GeV. The baryon mass has been put to 7 GeV, and the
wave function of light constituent quark in the system of quark-diquark has
been taken in accordance to the relation
$$
f = \sqrt{\frac{12}{M}} \Psi(0),
$$
so that in the $D$ meson we would have $f_D=200$ MeV. For the wave function of
diquark subsystem, we have used the estimates in the non-relativistic
model with the Buchm\" uller--Tye potential \cite{BT} with the color factor of
diquark. So, 
$$
\Psi^d(0) = 0.193\;\; {\rm GeV}^{3/2}.
$$
Further, it is quite evident that the estimates of spectator widths of free
heavy quarks do not depend on the system, wherein they are bound, so that we
can
exploit the results of the calculations performed earlier. We have chosen 
the quark
masses to be the same as in \cite{BB}, and we have put the corresponding
values, presented in table \ref{spectator}, as they stand in the paper
mentioned.
\begin{table}[th]
\begin{center}
\begin{tabular}{|c|c|c|c|}
\hline
mode     & $b\to c \bar u d$ & $b\to c \bar c s$ & $b\to c e^+ \nu$ \\
\hline
$\Gamma$ & 0.310 & 0.137 & 0.075 \\
\hline
mode     & $b\to c \tau^+ \nu$ & $c\to s \bar d u$ & $c\to s e^- \bar \nu$ \\
\hline
$\Gamma$ & 0.018 & 0.905 & 0.162 \\
\hline
\end{tabular}
\end{center}
\caption{The widths of inclusive spectator decays for the $b$- and $c$-quarks,
in ps$^{-1}$.}
\label{spectator}
\end{table}

Then the procedure, described above with the shown parameters, leads to the
lifetimes of the $\Xi_{bc}^{+}$ and $\Xi_{bc}^{0}$ baryons:
\begin{eqnarray}
\tau_{\Xi_{bc}^{+}} & = & 0.33\;\; {\rm ps},\\
\tau_{\Xi_{bc}^{0}} & = & 0.28\;\; {\rm ps}.
\end{eqnarray}
We can clearly see that the difference in the lifetimes caused by the decay
processes with the Pauli interference and weak scattering is about
20 \%. The relative contributions by various terms in the total
width of the baryons under consideration are presented in table \ref{br}.
\begin{table}[th]
\begin{center}
\begin{tabular}{|c|c|c|c|c|}
\hline
mode     & $\Gamma_b$ & $\Gamma_c$ & $\Gamma_{PI}$ & $\Gamma_{WS}$ \\
\hline
$\Xi_{bc}^+$ & 20 & 37 & 23 & 20\\
\hline
$\Xi_{bc}^0$ & 17 & 31 & 21 & 31\\
\hline
\end{tabular}
\end{center}
\caption{The branching fractions (in \%) of the various modes in the  
inclusive 
decays of the $\Xi_{bc}^{+}$ and $\Xi_{bc}^{0}$ baryons.}
\label{br}
\end{table}

Note, that the contributions by the Pauli interference and weak scattering,
depending on the baryon contents, can be significant - up to 
$40-50$ \%. The corrections due to the quark-gluon operators of dimension 5 are
numerically very small. The most important are the corrections due to the 
operator
of dimension 3, where the role of time dilation is essential.

For the semileptonic decays, whose relative fractions are presented in table
\ref{br-semi}, the largest corrections appear in the decays of $b$-quark
because of the Pauli interference, so that the corresponding widths practically
increase twice. This leads to the result the semileptonic widths of $b$- and
$c$-quarks in the electron mode are equal to each other, whereas for the 
spectator
decays, the width of the charmed quark is twice that of $b$.
\begin{table}[th]
\begin{center}
\begin{tabular}{|c|c|c|c|}
\hline
mode     & $\Gamma_á^{e\nu}$ & $\Gamma_b^{e\nu}$ & $\Gamma_b^{\tau\nu}$  \\
\hline
$\Xi_{bc}^+$ & 5.0 & 4.9 & 2.3\\
\hline
$\Xi_{bc}^0$ & 4.2 & 4.1 & 1.9\\
\hline
\end{tabular}
\end{center}
\caption{The branching ratios for the inclusive semileptonic widths of
$\Xi_{bc}^{+}$ and $\Xi_{bc}^{0}$, in \%.}
\label{br-semi}
\end{table}

As for the sign of terms, caused by the Pauli interference, it is basically
determined by the leading contribution, coming from the interference for the
charmed quark of the initial state with the charmed quark from 
the $b$-quark decay. In this way, the antisymmetric color structure 
of baryon wave function leads to the positive sign for the Pauli interference.

Finally, concerning the uncertainties of the estimates presented, 
we note that they are mainly related to the following:

1) the spectator width of charmed quark, where the error can reach 
50 \%, reflecting the agreement of theoretical evaluation with the lifetimes
of charmed hadrons, so that for the baryons under
consideration this term produces an uncertainty of $\delta \Gamma/\Gamma
\approx 10 \%$,

2) the effects of Pauli interference in the decays of beauty quark and in its
weak scattering off the charmed quark from the initial state, wherein we use
the non-relativistic wave function, which, being model-dependent, can 
lead to an
error estimated close to 30 \%, producing an uncertainty of $\delta
\Gamma/\Gamma \approx 15 \%$ in the total widths.

Thus, we estimate, that the uncertainty in the predictions of total widths for
the $\Xi_{bc}^{+}$ and $\Xi_{bc}^{0}$ baryons is about 25 \%.

\section{Conclusion}

In the present paper we have calculated the total lifetimes of baryons with two
heavy quarks of different flavors: $\Xi_{bc}^{+}$ and $\Xi_{bc}^{0}$, in the
framework of an Operator Product Expansion using the
inverse heavy quark mass technique. In this way, we have taken into account 
the QCD
corrections to the Wilson coefficients of the operators as
well as the mass corrections. The peculiar role is played by the four-quark
operators, responsible for the effects of Pauli interference and weak
scattering off constituents. These mechanisms are enforced due to the
two-particle phase space in the final state, so that these effects, depending
on the quark contents of baryons, provide the contribution close to 50 \%. For
the numerical estimates, we get:
\begin{eqnarray}
\tau_{\Xi_{bc}^{+}} & = & 0.33\pm 0.08\;\; {\rm ps},\\
\tau_{\Xi_{bc}^{0}} & = & 0.28\pm 0.07\;\; {\rm ps}.
\end{eqnarray}
We have also presented both the branching fractions of various contributions
into the total widths and the inclusive semileptonic decays.

This work is in part supported by the Russian Foundation for Basic Research,
grants 96-02-18216 and 96-15-96575. The work of A.I.Onishchenko was 
supported by International Center of Fundamental Physics in Moscow.

The authors would like to express the gratitude to Prof. A.Wagner and
members of DESY Theory Group for their kind hospitality during the visit 
to DESY, where this paper was written, as well as to Prof. S.S.Gershtein 
for discussions. The authors thank Prof. A.Ali for remarks, which 
essentially improved the presentation of paper.

\newpage
\setlength{\unitlength}{1mm}
\begin{figure}[p]
\vspace*{-9mm}
\begin{center}
\begin{picture}(400.,30.)
\hspace*{3cm}
\epsfxsize=9cm \epsfbox{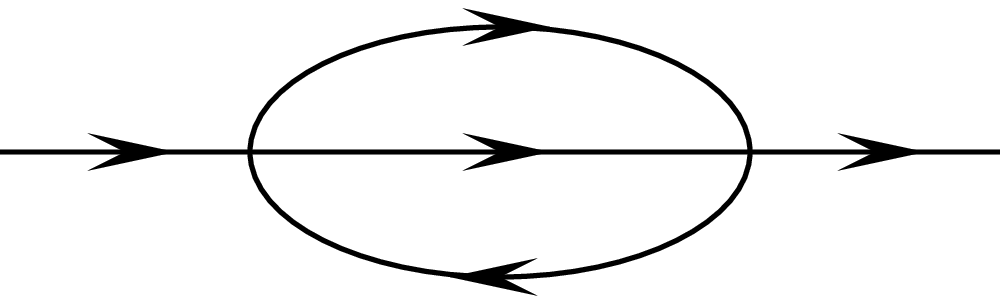}
\put(-82,24){$c$}
\put(-15,24){$c$}
\put(-50,24){$s$}
\put(-50,35){$u$, $l$}
\put(-50,3){$\bar d$, $\bar \nu$}

\end{picture}
\end{center}
\vspace*{-8mm}
\caption{The diagram of spectator contribution in the charmed quark decays.}
\label{fig1}
\end{figure}

\begin{figure}[p]
\vspace*{-9mm}
\begin{center}
\begin{picture}(400.,30.)
\hspace*{3cm}
\epsfxsize=9cm \epsfbox{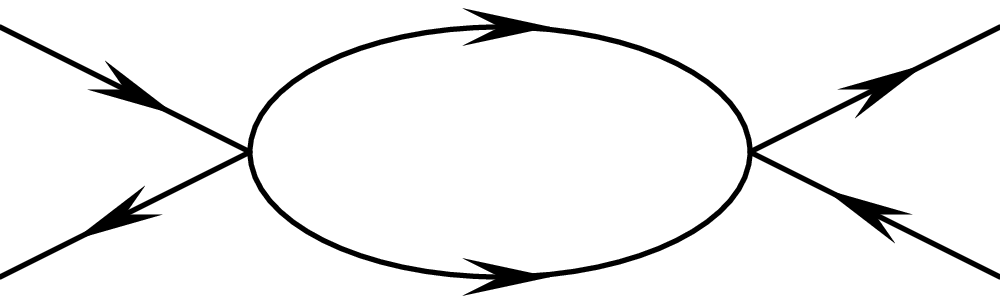}
\put(-82,30){$c$}
\put(-82,10){$u$}
\put(-15,30){$c$}
\put(-15,10){$u$}
\put(-50,35){$s$}
\put(-50,3){$\bar d$}

\end{picture}
\end{center}
\vspace*{-5mm}
\caption{The diagram for the contribution of Pauli interference in the decays
of charmed quark for the $\Xi_{bc}^{+}$ baryon.}
\label{fig2}
\end{figure}


\begin{thebibliography}{**}
\bibitem{HQET}
{\it M.Neubert}, Phys. Rep. 245 (1994) 259.
\bibitem{NRQCD}
{\it G.T.Bodwin, E.Braaten, G.P.Lepage}, Phys. Rev. D51 (1995) 1125,\\
Phys. Rev. D55 (1997) 5853.
\bibitem{BB}
{\it M.Beneke, G.Buchalla},  Phys. Rev. D53 (1996) 4991.
\bibitem{al}
{\it A.Ali, D.London}, Nucl. Phys. Proc. Suppl. {\bf 54A}, 297 (1997).
\bibitem{CDFbc}
{\it F.Abe et al.}, preprint FERMILAB-PUB-98-157-E. 1998. [hep-ex/9805034],
FERMILAB-PUB-98-121-E. 1998. [hep-ex/9804014].
\bibitem{thbc}
{\it S.S.Gershtein et al.}, preprint IHEP 98-22. 1998. [hep-ph/9803433];\\
{\it S.S.Gershtein et al.}, Uspekhi Fiz. Nauk. 165 (1995) 3.
\bibitem{ltcc}
{\it V.V.Kiselev, A.K.Likhoded, A.I.Onishchenko}, preprint DESY 98-079.
1998. [hep-ph/9807354].
\bibitem{spec}
{\it S.S.Gershtein, V.V.Kiselev, A.K.Likhoded, A.I.Onishchenko}, preprint
DESY 98-080. 1998. [hep-ph/9807375].
\bibitem{prod}
{\it A.V.Berezhnoy, V.V.Kiselev, A.K.Likhoded, A.I.Onishchenko}, Phys.
Rev. D57 (1997) 4385;\\
{\it D.Ebert, R.N.Faustov, V.O.Galkin, A.P.Martynenko, V.A.Saleev},
Z. Phys. C76 (1997) 111;\\
{\it J.G.K$\ddot o$rner, M.Kr$\ddot a$mer, D.Pirjol}, Prog. Part. Nucl.
Phys. 33 (1994) 787;\\
{\it R.Roncaglia, D.B.Lichtenberg, E.Predazzi}, Phys. Rev. D52 (1995) 1722;\\
{\it E.Bagan, M.Chabab, S.Narison}, Phys. Lett. B306 (1993) 350.
\bibitem{vs}
{\it M.B.Voloshin, M.A.Shifman}, Yad. Fiz. 41 (1985) 187;\\
{\it M.B.Voloshin, M.A.Shifman}, Zh. Exp. Teor. Fiz. 64 (1986) 698;\\
{\it M.B.Voloshin}, preprint TIP-MINN-96/4-T, UMN-TH-1425-96. 1996.
[hep-th/9604335].
\bibitem{bigi}
{\it I.Bigi et al.}, "B Decays", Second edition, ed. S. Stone (World
Scientific, Singapore, 1994)
\bibitem{4}
{\it G.P.Lepage et al.}, Phys. Rev. D46 (1992) 4052.
\bibitem{9}
{\it I.Bigi, N.Uraltsev, A.Vainshtein}, Phys. Lett. B293 (1992) 430,
Phys. Lett. B297 (1993) 477;\\
{\it B.Blok, M.Shifman}, Nucl. Phys. B399 (1993) 441, 459;\\
{\it I.Bigi et al.}, Phys. Lett. B323 (1994) 408.
\bibitem{10}
{\it A.V.Manohar, M.B.Wise}, Phys. Rev. D49 (1994)1310.
\bibitem{11}
{\it A.F.Falk et al.}, Phys. Lett. B326 (1994) 145;\\
{\it L.Koyrakh}, Phys. Rev. D49 (1994) 3379.
\bibitem{12}
{\it G.Altarelli et al.}, Nucl. Phys. B187 (1981) 461.
\bibitem{13}
{\it A.J.Buras, P.H.Weisz}, Nucl. Phys. B333 (1990) 66.
\bibitem{14}
{\it G.Buchalla}, Nucl. Phys. B391 (1993) 501.
\bibitem{15}
{\it Q.Hokim, X.Y.Pham}, Phys. Lett. B122 (1983) 297, Ann. Phys. 155 (1984)
202.
\bibitem{16}
{\it E.Bagan et al.}, Nucl. Phys. B432 (1994) 3, Phys. Lett.
B342 (1995)362;\\
{\it E.Bagan et al.}, Phys. Lett. B351 (1995) 546.
\bibitem{BT}
{\it W.Buchm\" uller and S.-H.H.Tye}, Phys. Rev. D24 (1981) 132.
\end{thebibliography}
\end{document}